%% file: rgmrMANUSCRIPT.tex
\documentclass[pre,showpacs,floatfix]{revtex4}

\usepackage{amsmath}
\usepackage{amssymb}
\usepackage{graphicx}
\usepackage{feynmf}
\usepackage[mathscr]{eucal}

\begin{document}

\title{Construction of coarse-grained order-parameters in non-equilibrium systems}

\date{\today}

\author{David E. Reynolds}

\affiliation{ICES, The University of Texas at Austin,
Austin, TX 78712}
\affiliation{Department of Physics and Institute for Genomic Biology, University of Illinois at Urbana-Champaign, 1110 West Green Street, Urbana, Illinois, 61801-3080, USA}

\begin{abstract}
We develop a renormalization group (RG) procedure that 
includes important system-specific features. 
The key ingredient is to systematize the coarse graining procedure 
that generates the RG flow.  The coarse graining technology
comes from control and operator theoretic model reduction.  
The resulting ``generalized'' RG is a consistent generalization of the Wilsonian RG.
We derive the form of the projection operator from the dynamics of a nonlinear wave 
equation and 
renormalize the distribution of initial conditions.  The probability density 
of the initial conditions is chosen to be the Boltzmann weight 
for a standard $\phi^4$-theory.  
In our calculation, we find that in contrast to conventional implementations of the RG,
na\"ive power counting breaks down.  
The RG-equations that we derive are different from those derived from the 
conventional RG.
\end{abstract}

\maketitle

\section{Introduction}
A recurring theme in the study of complex and biological systems is that the systems 
of interest are open.  The nature of the disturbances rendering the system open
may be stochastic, structured, or both.  Additionally, the stochastic nature of 
chemical networks, especially those with molecular-types in low abundance, ensures that
any intrinsic description 
must account for the structure and type of 
uncertainty.  Intuition from 
closed systems, 
with the possible exception of 
glassy systems \cite{cug97}, 
suggests that detailed questions are often too difficult to answer and that 
asymptotic questions are more amenable to theoretical treatments.
While for many open systems knowledge of the temporal asymptotics is both 
interesting and important, for others such knowledge may not carry any 
content or even exist.  However, just because 
temporal asymptotics may not exist for open systems,
this does not imply that 
there aren't robustly discernible features of open systems.  It simply means 
that the appropriate coarse-grained description(s) must be informed by 
the structure of the 
external and internal
disturbances and uncertainties.

A large class of open systems may be represented by equations of the form
\begin{equation}
\label{open_system1}
\frac{\partial\varphi(x,t)}{\partial t}  =f(\{\varphi\}) + u(x,t),
\end{equation}
where $\varphi$ and $u$ are vectors/functions in possibly infinite dimensional spaces.
 Without $u$, equation \eqref{open_system1}, represents a closed system or a model that describes 
a system with perfect certainty.  With $u$, the system is inherently open subject to additive
uncertainty.
$u$ represents  generic driving as well as a possible noise source to which 
the system is exposed.  If $u$ is generated by a continuous stochastic process, this
 imposes a specific structure on the noise.  Similarly, constraining $u$ to belong to a particular
function space but remaining otherwise arbitrary also imposes a specific structure on the noise.
The system, as described, may be mapped to a field theory via generalized 
Martin-Siggia-Rose (MSR)/closed-time-path (CTP) methods~\cite{martin73,cooper01, zanella02, zhou85}.
In particular, the path-integral representation for 
the probability for the system to be in state $\phi$ at time $t$ given that the initial state was 
 $\phi_0$ at time $t_0$ is given by \cite{zinn_justin96}
\begin{equation}
P(\phi, t|\phi_0, t_0)  = \int  D\varphi\, \exp\{-\frac{1}
{2\Delta x}\int_{t_0}^{t} d\tau\partial_{\varphi} f \} \,
\delta [ u - \partial_{\tau}{\varphi} +  f(\varphi) + \phi_0\delta(\tau-t_0) ] \,,
\label{eqn:c1_1pi}
\end{equation}
When the initial state is only known probabilistically, then the probability for the 
system to be in state $\phi$ at time $t$ is given by
\begin{equation}
P(\phi,t) = \int D\phi_0 P(\phi_0,t_0)P(\phi, t|\phi_0, t_0)  
\label{eqn:c2_1pi}
\end{equation}
We introduce this representation because later we will explicitly utilize it in the context of applying the renormalization group (RG) to 
the initial conditions. 
It is important to note that upon combining Equations \eqref{eqn:c1_1pi} and \eqref{eqn:c2_1pi}, 
integrating out $\varphi$ does nothing to dress the distribution $P(\phi_0,t_0)$.  This simply 
reflects the fact that the initial conditions are imposed as a constraint.  On the other hand,
integrating out $\phi_0$ would then add stochastic contributions to Equation \eqref{open_system1}.

Suppose $u$ is an arbitrary input to the system that is square-integrable in time. We consider 
the states or regions in phase space that are most accessible 
via driving to be responsible
for describing the essential characteristics of the system.  
This is analogous to the energy landscape picture in statistical mechanics where 
fluctuations govern which states
contribute the most to the statistics of the system.
We use this inherently open systems perspective of the importance of states 
to specify how to coarse grain and, consequently, to generate RG equations.
The key step in generalizing the RG lies in ascertaining how to coarse grain.

Although it is already known that the RG is not 
a black box routine, the purpose of this work is to make it more algorithmic.
  It is easy to be misled into thinking that
the  RG already is algorithmic
because its key ingredients are coarse graining 
and rescaling the system variables~\cite{wilson74,wilson75,goldenfeld92,shankar94}.
However, fully algorithmic implementations of the RG fail 
for large classes of problems because 
it is not possible to ignore the physics of a system and 
expect to obtain meaningful results.  
Capturing the essential physics requires isolating the appropriate models
and the structure of perturbations and uncertainties.
Consequently, this process is system-specific.
Additionally, the scale on which the physics is observed must 
be specified.
For instance, for bosonic theories  
the long-wavelength physics is investigated,
while for fermionic system it is the physics near the Fermi surface. 
These considerations suggest that the primary obstacles to automation 
are the model identification and coarse graining processes.
In this article we present a RG procedure
that accounts for these system-specific obstacles  
and apply it to renormalize the initial conditions of a nonlinear wave equation.
We specifically focus on a nonlinear wave equation as opposed to a reaction 
diffusion equation because the extra time derivative increases the number of 
Green's functions that can contribute to the dynamics of the system.  
As we will see, this can have a significant impact on coarse graining.
We do not renormalize the dynamical equations because
the structure of the dynamical action renders the perturbative loop corrections
uninteresting.  The general form of how projection operators, 
generating the RG, act on the dynamical fields is discussed in \cite{degenhard02}.

The obstacles mentioned above
arise from the (mis)identification of observables.
We use techniques taken from operator theoretic interpolation theory and control theory 
to systematically identify observables that respect the system dynamics.
The conventional implementation of the renormalization group relies upon 
discerning equilibrium observables at predetermined scales.  
Coarse graining nonequilibrium systems must respect the nature of the dynamics and the structure of 
the uncertainty.  Emerging fields like systems biology are producing many new problems that are begging
for 
such a treatment \cite{paulsson03,paulsson04}.  
This work provides a step towards the development of a
framework that will be applicable to such systems.

While equation \eqref{open_system1} addresses the effects 
of perturbations to the system, it does not account for the 
consideration that it may not be desirable to 
describe every 
observable.  For instance, complete characterization of a system over 
all scales may carry significant computational overhead.  Furthermore, due
to experimental constraints, it may not be possible to measure all observables
either.  
Consequently, it is beneficial to explicitly include 
the possibility of multiscale or constrained observation.
One of the simplest cases is when the observables are linearly dependent on the 
$\varphi$-observables.  The description of such open systems 
takes the form
\begin{eqnarray}
\label{open_system2}
\dot{\varphi} =f(\varphi) + Bu, \nonumber \\
\psi = C\varphi,
\end{eqnarray}
where $\psi$ reflects that only some subspace or, more generally, subset
of phase space is directly measurable.  The operators $B$ and 
$C$ respectively specify the structure of how noise may enter the system and 
which states are considered as observables.  More generally the observables may 
be nonlinear functions or functionals of $\varphi$, however without mapping the 
problem to a yet higher dimensional description, that complication is beyond the scope of
the analysis presented herein.
Considering coarser observables $\psi$
influences the relative importance of the internal states, $\varphi$, 
because many of them simply either do not or only marginally contribute to the 
physical observables $\psi$.  Hence, the choice of observables can strongly influence 
how models describing a system ought to be coarse grained.   
For instance, in the original analysis of Feynman
and Vernon \cite{feynman63} or Caldeira and Leggett \cite{caldeira83} 
characterizing a particle in a heat bath (albeit in a quantum mechanical context), 
the microscopic degrees of freedom
contributing to the heat bath were systematically projected away thereby leaving their effective 
influence on the particle.  Implicitly, in their analysis a choice of $C$ was made and 
consequently led to the derivation of effective stochastic equations describing 
the physics of the chosen observables.

The outline of the paper is as follows.  In Section \ref{cg_balance} we introduce 
the coarse graining procedure.  We use the procedure in Section \ref{projOP} to identify
the projection operator for diffusion and wave equations.  
In Section \ref{RG_analysis} the projection operators 
are used to generate RG-flows.  We derive the associated RG-equations and 
contrast the RG-flows.  Our conclusions are given in Section \ref{conclusion}.
Two Appendices provide detailed calculations related to the derivation of the 
projection operators and the RG-equations.

\section{Coarse graining via balancing}
\label{cg_balance}
We coarse grain by 
retaining the states contributing most to the dynamical response of the system.
While technically different, this work is in same spirit as that by 
Chen et al \cite{chen96} and more recently by Degenhard and Rodr\'iguez \cite{degenhard02,degenhard05}.
As a starting point, we coarse grain equation \eqref{open_system2} from its 
linearization about a particular solution with $u = 0$.  The reasoning for this approach 
follows many of the same arguments why linear-response theory has been used with such frequency.
Specifically some of those reasons are that 1) the approach is amenable to calculation, 
2) it is often known that the system operates about a particular equilibrium, 
and 3) for large-amplitude noise, the linear-response dominates since the system then does not 
get trapped in local equilibria.  
As a simplification, we only consider linearizations about dynamical 
steady states.  
In systems biology, similar approximations are ubiquitously used when describing biochemical 
networks \cite{paulsson03}.
The linearizations are generically described by  
\begin{eqnarray}
\label{open_system4}
\dot{\tilde{\varphi}} = A\tilde{\varphi}+ Bu, \nonumber \\
\tilde{\psi} = C\tilde{\varphi},
\end{eqnarray}
where $B$ and $C$ are the same operators as those in \eqref{open_system2}, $\varphi = \varphi_{eq} + \tilde{\varphi}$ 
where $\varphi_{eq}$ is an equilibrium solution,
and $A = \partial_{\varphi}f|_{\varphi_{eq}}$.
Associated with 
equation \eqref{open_system4} are 
invariants known as Hankel singular values (HSV's)~\cite{glover84,peller03}.  
These invariants have many nice properties, but 
their most significant being that they give explicit information about the Green's functions 
(i.e.~dynamical evolution operator) for the system.
These invariants are most easily calculated by solving for 
positive operators $X$ and $Y$, 
that are determined by the equations
\begin{eqnarray}
\label{Lyap1}
\frac{\mathrm{d}X}{\mathrm{d}t_f} = AX+XA^{\dagger}+ \tilde{B}\tilde{B}^{\dagger}; \ \ X(0)=0, \\
\label{Lyap2}
\frac{\mathrm{d}Y}{\mathrm{d}t_f} = A^{\dagger}Y+YA + C^{\dagger}C; \ \ Y(0) = 0.
\end{eqnarray}
If we define another operator $W$ by 
\begin{equation}
\label{HSV}
W^2 = XY,
\end{equation}
then the HSV are nonnegative real numbers, $\sigma_{\max} \ge \sigma_{\kappa} \ge \sigma_{\min}$ 
that comprise the spectrum of the operator $W$.  
HSV's provide a precise measure 
of the error incurred by approximating 
the effect $u$ has on $\tilde{\psi}$ with reduced order models.
The HSV's may be interpreted as supplying a 
measure of the importance of the internal states $\tilde{\varphi}$.  
If $W$ is invertible, it is always possible to find a coordinate system,
called {\it balanced coordinates}, such that 
$X=Y=\mathrm{diag}(\sigma_{\max},\hdots,\sigma_{\min})$.
When equation \eqref{open_system4} is transformed to balanced coordinates, 
the best reductions are those that project out the states corresponding 
to small HSV.  In other words, the ordering of the HSV, at least locally around
an equilibrium configuration in phase space, specifies how to coarse grain a system. 
An in depth treatment of this material may be found in \cite{dullerud00}.
It is also sometimes possible to ``balance'' the full nonlinear system~\cite{scherpen93}.

The RG can easily be adapted for HSV-based coarse graining.
Operator theoretic approaches to the RG \cite{muller96,bach98,bach03} demonstrate that 
coarse graining in the Wilsonian RG is equivalent to multiplying operators or states by projection
operators.
The essence of this work is to use HSV's to \emph{identify} the projection operator. 
As before, suppose that ${\bf \kappa}$ is a vector index that orders the HSV's $\sigma_{\bf \kappa}$
for equation \eqref{open_system4} from largest to smallest.
A generalized Wilsonian RG procedure is obtained by:
1) transforming both the initial conditions and the dynamical variables 
to balanced coordinates,
\begin{equation}
\label{balancing}
\hat{\varphi}({\bf \kappa},t) = \int R({\bf \kappa},{\bf x})\varphi({\bf x},t) \mathrm{d}{\bf x},
\end{equation}
so that the distribution $P(\phi,t)$ 
takes the form,
\begin{eqnarray}
\label{part_fnc1}
P(\phi,t) = \int D\phi_0 D\varphi D\bar{\varphi}
\exp\left(-S_{dyn}\left({\bf g},\{\varphi\},\{\bar{\varphi}\},\{\phi_0\}\right) -S\left({\bf g},\{\phi_0\}\right) \right) \nonumber \\ 
= \int D\hat{\phi}_0 D\hat{\varphi} D\hat{\bar{\varphi}}\mathcal{J}\exp\left(-S_{dyn}\left(\tilde{\bf g},\{\hat{\varphi}\},\{\hat{\bar{\varphi}}\},\{\hat{\phi}_0\}\right) -S\left(\tilde{\bf g},\{\hat{\phi}_0\}\right)\right),
\end{eqnarray}
where $S_{dyn}$ is the action representing the dynamics, $S$ is the action of the initial conditions, ${\bf g}$ is the original set of coupling constants/functions, $\bar{\varphi}$ is the dual field
that arises from representing the functional-$\delta$ in equation \eqref{eqn:c1_1pi}
as an exponential, $\mathcal{J}$ is the Jacobian 
from equation \eqref{balancing}, and $\tilde{\bf g}$ is the resulting transformed set of coupling constants;
2) integrating out ${\bf \kappa}$-shells about $\sigma_{\min}$ analogously to wave-vector shells;
and 3) rescaling ${\bf \kappa}$ and $\hat{\bf \phi}$ appropriately.  
An interesting but technically challenging variant of this procedure 
is to integrate out $\sigma_{\bf \kappa}$-shells instead of ${\bf \kappa}$-shells about $\sigma_{\min}$.
Integrating out a single  $\sigma_{\bf \kappa}$-shell
may entail integrating out an entire subspace in ${\bf x}$-space because
$\sigma_{\bf \kappa}$
does not necessarily respect spatial dimension.  
The technical challenge lies in rescaling  $\sigma_{\bf \kappa}$.  It is not clear that rescaling 
 $\sigma_{\bf \kappa}$ will recover the full ${\bf \kappa}$-space thereby 
 generating a meaningful RG.

A standard interpretations of the large wave-vector cutoff in the RG is that the inverse-cutoff
 is proportional to the 
smallest length scale of the system~\cite{wilson74,wilson75}.  
The idea of projecting out the large wave-vector physics down 
to the small wave-vector physics has the interpretation of homogenizing the system to its continuum, large-scale
limit.  However, interfaces that are smooth on short spatial-scales may actually appear sharp when viewed
on large scales.  Representing such  effective, sharp interfaces, in terms of Fourier (wave-vector) 
modes then requires a huge number of modes (the largest mode being the cutoff).
Other representations, like 
wavelet representations, are much better equipped to describe the singular nature of such interfaces.  
Mathematically this means that the cutoff is specifically chosen because it provides us with a 
starting point for coarse-graining and an ordering relation that determines in which direction to coarse grain.  
The cutoff can be thought of entirely in  approximation/interpolation theoretic terms.
The initiation of coarse graining starts with modes contributing the least to the 
phenomena of interest.  The ordering-relation results from the ordering of
modes by their contribution to the approximation.
$\sigma_{\min}$ is a 
 generalization of the standard wave-vector cutoff and the ordering-relation, determined by the ordering
of the HSVs ($\sigma_{\bf \kappa}$), is the generalization of standard scale-ordering (small scale to 
large scale).  
The RG has been studied in the context of approximation/interpolation theory before.  
RG methods have been related to theory of Pad\'e approximants by Baker \cite{baker96}, 
to wavelet theory by Battle \cite{battle87,battle88}, and to 
subspace/Lanzcos methods by White \cite{white92,white93}.

\section{Identification of the projection operator}
\label{projOP}
Here we apply the aforementioned coarse graining procedure to identify the 
projection operator for linear diffusion and linear and nonlinear wave equations.  Since 
the procedure requires us to expand about a dynamical steady state, if we expand about 
the trivial solution, then the nonlinearity does not change anything.  In this section we 
find that by considering \emph{slow} observables for the diffusion equation and wave 
equation, projecting out large wave-vector shells is appropriate.  When 
we additionally consider the fast, kinetic observables for the wave-equation, we arrive 
at quite a different result.
This results from the extra time derivative of the wave-equation and is the very reason why 
we consider the nonlinear wave-equation in this paper.

\subsection{Linear diffusion equation}
We first consider the (driven) diffusion equation
\begin{equation}
\label{diffusion}
\partial_t \phi = D\nabla^2\phi + \gamma u.
\end{equation}
In this example, $B=\gamma$, $C=1$, $A = D\nabla^2$ and we take $t_f \to \infty$.
By considering a stable system over an infinite time horizon, we only need to solve the Lyapunov equations,
\begin{eqnarray}
\label{Lyap3}
AX+XA^{\dagger}+ \tilde{B}\tilde{B}^{\dagger} = 0,\\
\label{Lyap4}
A^{\dagger}Y+YA +C^{\dagger}C = 0,
\end{eqnarray}
instead of equations \eqref{Lyap1}-\eqref{Lyap2}.
By taking the Fourier transform,
equations \eqref{Lyap3}-\eqref{Lyap4} become
\begin{eqnarray}
\label{Lyap3b}
2D|{\bf k}|^2X_{\bf k}+ |\gamma|^2 = 0,\\
\label{Lyap4b}
2D|{\bf k}|^2Y_{\bf k} + 1 = 0.
\end{eqnarray}
From equation \eqref{HSV}, it follows 
that in balanced-coordinates $W$ is given by
\begin{equation}
\label{diff_gramian}
W_{\bf k} = \frac{|\gamma|}{2D|{\bf k}|^2}.
\end {equation}
Here the index, ${\bf \kappa}$ for the HSV's is just $|{\bf k}|$.  Thus, for the diffusion
equation, the most important states are those that correspond to small wave-vector.  Thus,
local coarse graining is appropriate because the smallest observable ``fluctuations''
are due to the short-wavelength physics.
The smallest error is incurred by projecting out 
large wave-vectors.  We have thus identified the projection operator for 
the diffusion equation and it \emph{exactly} coincides with what we would expect.
Diffusive dynamics spatially homogenizes disturbances, thus, intuitively we 
already know that local coarse graining is desirable.

\subsection{Linear wave equation}
As a second example, we consider the (driven) linear wave equation
\begin{eqnarray}
\label{linear_wave_equation1}
\partial_t^2 \phi = v^2\nabla^2\phi + \gamma u, \nonumber \\
{\bf y} = \phi. 
\end{eqnarray}
By the units of $u$, it represents a true force acting on $\phi$.  This, in addition to 
the fact that $\phi$ is the ``measurable'' quantity, implies that we have 
isolated our attention on $\phi$-based observables.  This choice is 
strongly influenced by equilibrium statistical mechanics and thermodynamics. 
We have completely neglected $\pi$, the field 
conjugate to $\phi$, that represents the kinetic contributions to the system.
When posed as a set of first order equations, equation \eqref{linear_wave_equation1} becomes
\begin{equation}
\label{linear_wave_equation2}
\left[\begin{array}{c}
\partial_t \phi  \\ \partial_t \pi \\ {\bf y}  
\end{array}\right] =
\left[\begin{array}{ccc}
0 & 1 & 0 \\
v^2 \nabla^2 & 0 & \gamma \\
1 & 0 & 0 
\end{array}\right]
\left[ \begin{array}{c}
\phi \\ \pi \\ u
\end{array}\right]
\end{equation}
By smoothing out the time-cutoff at $t_f$ with a damped exponential
in the integral representation of the solution of equations \eqref{Lyap1} and \eqref{Lyap2}
the problem simplifies to solving Lyapunov equations.  This smoothing process is also 
known as exponential discounting.  With the given form of $B$ and $C$ in 
this problem, we find that the matrix of HSV's, $W$, is approximately given by
\begin{equation}
\label{lwe_gramian}
W^{bal}_{\bf k} \approx \frac{|\gamma|}{4av|{\bf k}|}\otimes I_{2\times2},
\end {equation}
where $I_{2\times 2}$ is the $2\times 2$ matrix identity, $a \sim 1/t_f$, 
and $\otimes$ is the tensor (dyadic) product. 
The calculation that produces this result is a special case of the calculation in Appendix \ref{gram_der}.
As with the diffusion equation, short-wavelength physics does not significantly contribute to 
the response, so locally coarse graining is appropriate. 
In the examples considered here, the dynamical equations that specify the projection operator are linear.
Thus, no approximation has been made and we expect that the corresponding projection operators 
are globally valid in phase space.

\subsection{Nonlinear wave equation with nonequilibrium observables}
We now consider a nonlinear wave equation with a nonequilibrium set of observables.  
As will be seen, a surprising result is that the choice of observables forces us to 
nonlocally coarse grain.  The nonlocality of the coarse graining has very interesting implications with 
regard to the resulting induced RG flow.  
The (driven) equations of motion that we are considering are
\begin{eqnarray}
\partial_t \phi = \pi + \alpha_1 u_1 \nonumber \\
\partial_t \pi = \nabla^2\phi + \frac{\lambda}{3!}\phi^3 + \alpha_2 u_2 \nonumber \\
{\bf y} = \left[\begin{array}{c} \beta_1\phi \\ \beta_2\pi \end{array}\right],
\end{eqnarray}
where $\phi$ and $\pi$ are real-valued fields.
The driving now includes generalized forces in addition to ``true'' forces.
By expanding around equilibrium solutions of $\nabla^2\phi=0$ we find that 
for each real-space position ${\bf x}$, 
\begin{eqnarray}
\label{BnC}
B = \left(\begin{array}{cc} \alpha_1 & 0 \\ 0 & \alpha_2 \end{array}\right) \ \textrm{and} \\ 
C = \left(\begin{array}{cc} \beta_1 & 0 \\ 0 & \beta_2 \end{array}\right).
\end{eqnarray}
This driving allows for more states in $(\phi,\pi)$-phase space
to be accessible compared to the driving in equation \eqref{linear_wave_equation2}.  
This, in combination with the form of ${\bf y}$,
ensures that both $\phi$ and $\pi$-dependent observables are being considered.
By using exponential discounting, we find that the 
diagonal operator of HSV's is given by (see Appendix \ref{gram_der})
\begin{eqnarray}
\label{nlwe_gramian}
W_{\bf k} \approx \frac{1}{4a}\left[(\alpha_2^2|{\bf k}|^{-1}+\alpha_1^2|{\bf k}|)\right. \nonumber \\
\left.\times(\beta_1^2|{\bf k}|^{-1}+\beta_2^2|{\bf k}|)\right]^{1/2}\otimes I_{2\times2}.
\end{eqnarray}
$W_{\bf k}$ does not have the HSV's ordered from largest to smallest, so it is 
not truly in balanced coordinates.  It is immediately apparent that the HSV's 
are large for both large and small magnitude wave-vector.  
A heuristic explanation for this strange result is that for large wave-vector,
$\pi$ is a pathologically ``fast'' variable.  However, by driving $\pi$ with $u_1$
over all wave-vector, this permits the fast resonances to be excited at large wave-vector.  
The pathological nature of $\pi$ as an observable
is analogous to the pathological nature of considering $\dot{\xi}$ an observable where
$\xi$ satisfies a Langevin equation~\footnote{$\dot{\xi}$ is pathologically fast compared to $\xi$}. 
In this sense, $\pi$ is a nonequilibrium observable of sorts.
Furthermore, because both the small and large wavelength 
physics contributes strongly to the response of the system, local coarse graining cannot be correct.
The appropriate coarse graining is nonlocal. 

\section{RG analysis}
\label{RG_analysis}
Now that we have specified how to coarse grain, we must specify the distribution of initial conditions that 
we wish to coarse-grain.  Should we choose a Gaussian distribution of initial conditions, then it turns out
that provided that the action, $S\left({\bf g},\{\phi_0\}\right)$, 
does not have a non-local spatial dependence, then the integration regions
decouple in Fourier space and the RG is trivial.  We thus consider a distribution of initial 
conditions that is local and quartic.  
Specifically, we consider the action for the initial conditions to be 
\begin{displaymath}
S({\bf g},\{\phi_0\}) =  \int \mathrm{d}{\bf x}\left( \frac{1}{2}\nabla\phi_0\nabla\phi_0 + 
  \frac{\lambda}{4!}\phi_0^4\right).
\end{displaymath}
It is important to remark here that we consider the gradient term to be local (or marginally local).
Equivalently, in Fourier space, the action is given by 
\begin{eqnarray}
\label{action}
S({\bf g},\{\hat{\phi}_0\}) =  \frac{1}{2(2\pi)^{D}}\int \mathrm{d}{\bf k} |{\bf k}|^2|\hat{\phi}_0({\bf k})|^2  
+ \nonumber \\
  \frac{\lambda}{4!} \int \prod_{n=1}^4 \frac{\mathrm{d}{\bf k}_n}{(2\pi)^D}  
\delta\left(\sum_{j=1}^4 {\bf k}_j\right) 
\hat{\phi}_0({\bf k}_1)\hat{\phi}_0({\bf k}_2)\hat{\phi}_0({\bf k}_3)\hat{\phi}_0({\bf k}_4), 
\end{eqnarray}
where $D$ is the spatial dimension of the system.
Rather than being wholly unmotivated, with the addition of kinetic terms,
the CTP-method \cite{zhou85,zanella02} may be used to derive 
a dynamical action from the above $S$ that is approximately the same as $S_{dyn}$ in Equation \eqref{part_fnc1}.
Also, in the remainder, to avoid unnecessary subscripts, we denote $\hat{\phi}_0$ by $\hat{\phi}$
with the understanding that we're performing the RG on the distribution of initial conditions.

\subsection{RG equations from local coarse graining}
Here we introduce a large wave-vector cutoff $\Omega$.
We define 
$\hat{\phi} = \hat{\phi}_< + \hat{\phi}_>$, where 
$\hat{\phi}_<$ is only nonzero for $|{\bf k}| \le \Omega/b$ ($b>1$)and
$\hat{\phi}_>$ is only nonzero for $\Omega/b < |{\bf k}| < \Omega$.
Integrating out 
wave-vector shells between  $\Omega/b$ and  $\Omega$ entails integrating 
out the $\hat{\phi}_>$ fields.  We then rescale by defining
\begin{equation}
\hat{\phi}_<({\bf k}) = Z\varphi_1(b{\bf k})
\end{equation}
and ${\bf p} = b{\bf k}$.  
Although we do not start with a ``mass'' term in the action (i.e. $m^2\phi^2$), such a term
is generated by the RG flow.  
Upon following the Wilson RG variant of what is described in 
Appendix \ref{rgeq_derivation} or the analysis done by Shankar in \cite{shankar94}, we obtain the RG equations
\begin{eqnarray}
\label{oldRG1}
\partial_{l}\bar{\lambda} = (4-D)\bar{\lambda} - \frac{3}{2}\frac{S_D}{(2\pi)^D}\bar{\lambda}^2 +\mathscr{O}(\bar{\lambda}^3) 
\\
\label{oldRG2}
\partial_{l}\bar{m}^2 = 2\bar{m}^2 + \frac{S_D}{2(2\pi)^D}\bar{\lambda}(1-\bar{m}^2) +\mathscr{O}(\bar{\lambda}^2).
\end{eqnarray}
$\bar{\lambda}$ and $\bar{m}^2$ are dimensionless and result from appropriately 
rescaling $\lambda$ and $m^2$ by $\Omega$.
The RG flow and fixed points for these equations are well studied. Two fixed points are the Gaussian 
fixed-point, $\bar{m}^2 = 0$ and $\bar{\lambda} = 0$, and the Wilson-Fisher fixed-point, 
 $\bar{m}^2 \approx -(4-D)/6$ and $\bar{\lambda} \approx 2(4-D)(2\pi)^D/(3S_D)$.  Following standard 
convention, we approximate the Wilson-Fisher fixed-point in powers of $4-D$
(the $\epsilon$-expansion).  
We provide the form of these RG equations because, as will soon be evident, they differ 
greatly from those that we derive in the next section.

\subsection{RG equations from nonlocal coarse graining}
In the case where $\phi$ and $\pi$ are treated on equal footing as observables, which in general may 
not be the case, $\alpha_1=\alpha_2=\alpha$ and $\beta_1=\beta_2=\beta$.  In the remainder, we treat 
this particular case.  Furthermore, without loss of generality, we set $\alpha = \beta = 1$.
In this case, equation \eqref{nlwe_gramian} indicates that 
 the $|{\bf k}|=1$ states are the least important.  Thus, for the purposes of the RG, the $|{\bf k}|=1$
hyper-surface serves as our analog of the wave-vector cutoff.  Implementing the second step of 
the procedure for generalized RG involves integrating out ${\bf k}$-shells {\it away} from the
$|{\bf k}|=1$ surface.  Rather than transform the system into the balanced ${\bf \kappa}$-coordinates,
out of convenience, we coarse grain the system in ${\bf k}$-space.  

In order to coarse grain, we let $\hat{\phi} = \hat{\phi}_< + \hat{\phi}_m + \hat{\phi}_>$ where 
$\hat{\phi}_<$ is only nonzero for $|{\bf k}| \le \Lambda$,
$\hat{\phi}_m$ is only nonzero for $\Lambda < |{\bf k}| < \Lambda^{-1}$, 
and $\hat{\phi}_>$ is only nonzero for $\ |{\bf k}| \ge \Lambda^{-1}$, where $\Lambda<1$.
Using this decomposition, the path integral measure factors as 
$\mathscr{D}\hat{\phi}=\mathscr{D}\hat{\phi}_<\mathscr{D}\hat{\phi}_m\mathscr{D}\hat{\phi}_>$.
The RG equations are then induced by integrating out $\hat{\phi}_m$ and then rescaling 
the wave-vectors and fields.  For this problem, the rescaling procedure requires that 
\begin{eqnarray}
\label{rescale}
\hat{\phi}_<({\bf k}) = Z_<\varphi_1(\Lambda^{-1}{\bf k}),  \\
\hat{\phi}_>({\bf k}) = Z_>\varphi_2(\Lambda{\bf k}),
\end{eqnarray}
and ${\bf p} = \Lambda^{-1}{\bf k}$ for $|{\bf k}| \le \Lambda$ and
 ${\bf p} = \Lambda{\bf k}$ for $|{\bf k}| \ge \Lambda^{-1}$. Na\"ive power
counting breaks down as a direct result of rescaling in the two disjoint wave-vector 
regimes.

Although we start with a theory where ${\bf g} = (1,\lambda,0,\hdots)$ we can 
expect that the RG transformations may generate new nonlinear terms 
and that the coupling constants may become coupling functions.  
In fact, ${\bf g}$ flows towards having an 
infinite number of nontrivial components.  
In particular, in a complete treatment, the coupling constant $\lambda$ becomes a coupling function,
$\lambda({\bf p_1},{\bf p_2},{\bf p_3},{\bf p_4})$.  However, if we focus purely
on the constant contributions of $\lambda({\bf p_1},{\bf p_2},{\bf p_3},{\bf p_4})$ in the
different wave-vector scaling regimes, we see that it may be decomposed 
into the couplings $\{\lambda_{(i,4-i)}\}_{i=1}^4$.  Here $\lambda_{(i,j)}$ 
represents the coupling constant in the scaling-regime with $i$ wave-vectors having $|{\bf p}_n|<1$
and $j$ wave-vectors having $|{\bf p}_n|>1$.
We denote the mass terms for $|{\bf q}|<1$  and  $|{\bf q}|>1$ 
respectively by $m^2_<$ and $m^2_>$. 
If we let $\Lambda = e^{-dl}$, then to first loop order the RG equations for 
$\{\lambda_{(i,4-i)}\}_{i=1}^4$ are
\begin{eqnarray}
\label{RG_eqns1}
\partial_l\lambda_{(4,0)} = (4-D)\lambda_{(4,0)} - \frac{3}{2}\frac{S_D}{(2\pi)^D}
\left[\lambda_{(4,0)}^2(m^2_<+1)^{-2} + \right.\nonumber \\
\left. + 2\lambda_{(3,1)}^2(m^2_<+1)^{-1}(m^2_>+1)^{-1} + \lambda_{(2,2)}^2(m^2_>+1)^{-2}\right] \\
\partial_l\lambda_{(0,4)} = (D-4)\lambda_{(0,4)} - \frac{3}{2}\frac{S_D}{(2\pi)^D}
\left[\lambda_{(2,2)}^2(m^2_<+1)^{-2} + \right.\nonumber \\
\left. + 2\lambda_{(1,3)}^2(m^2_<+1)^{-1}(m^2_>+1)^{-1} + \lambda_{(0,4)}^2(m^2_>+1)^{-2}\right]
\end{eqnarray}
\begin{eqnarray}
\partial_l\lambda_{(1,3)} = -2\lambda_{(1,3)} - 6\frac{S_D}{(2\pi)^D}
\left[\lambda_{(2,2)}\lambda_{(3,1)}(m^2_<+1)^{-2} + \right.\nonumber \\
\left. + 2\lambda_{(1,3)}\lambda_{(2,2)}(m^2_<+1)^{-1}(m^2_>+1)^{-1} + \lambda_{(0,4)}\lambda_{(1,3)}(m^2_>+1)^{-2}\right] \\
\partial_l\lambda_{(3,1)} = -2(D-1)\lambda_{(3,1)} - 6\frac{S_D}{(2\pi)^D}
\left[\lambda_{(4,0)}\lambda_{(3,1)}(m^2_<+1)^{-2} + \right.\nonumber \\
\left. + 2\lambda_{(3,1)}\lambda_{(2,2)}(m^2_<+1)^{-1}(m^2_>+1)^{-1} + \lambda_{(2,2)}\lambda_{(1,3)}(m^2_>+1)^{-2}\right] 
\end{eqnarray}
\begin{eqnarray}
\partial_l\lambda_{(2,2)} = -D\lambda_{(2,2)} - 3\frac{S_D}{(2\pi)^D}
\left[\left(\lambda_{(2,2)}\lambda_{(4,0)} + 2\lambda_{(3,1)}^2\right)(m^2_<+1)^{-2} + \right.\nonumber \\
\label{RG_eqns1e}
\left. + 2\left(\lambda_{(1,3)}\lambda_{(3,1)}+2\lambda_{(2,2)}^2\right)(m^2_<+1)^{-1}(m^2_>+1)^{-1} + \left(\lambda_{(0,4)}\lambda_{(2,2)}+2\lambda_{(1,3)}^2\right)(m^2_>+1)^{-2}\right] 
\end{eqnarray}
The RG equations for the mass 
terms are 
\begin{eqnarray}
\label{RG_eqns2a}
\partial_lm^2_< = 2m^2_< + \frac{S_D}{2(2\pi)^{D}}\left(\frac{\lambda_{(4,0)}}{m_<^2+1}+\frac{\lambda_{(2,2)}}{m_>^2+1}\right)+\mathscr{O}({\lambda^2}) \\
\label{RG_eqns2b}
\partial_lm^2_> = -2m^2_> + \frac{S_D}{2(2\pi)^{D}}\left(\frac{\lambda_{(2,2)}}{m_<^2+1} +\frac{\lambda_{(0,4)}}{m_>^2+1}\right)+\mathscr{O}({\lambda^2})
\end{eqnarray}

The first thing to notice in equations \eqref{RG_eqns1}-\eqref{RG_eqns1e} 
is that the contributions from tree level,
the linear terms, indicate that the couplings  involving a mixing of wave-vectors 
(i.e.~ $i,j\ne 0 $) are irrelevant.  
The terms that are relevant, appear similar to equations \eqref{oldRG1}-\eqref{oldRG2}.
In fact, if we expand about $m_<^2=0$, make the identifications $\lambda_{(4,0)}=\bar{\lambda}$
and $m_<^2 = \bar{m}^2$, and set the rest of the coupling constants to zero, then 
Equations \eqref{RG_eqns1} and \eqref{RG_eqns2a} are exactly the same as 
equations \eqref{oldRG1}-\eqref{oldRG2}. 
Thus, if it were possible to ignore all other relevant perturbations, then
the new set of RG equations would retain the 
standard fixed points (Gaussian and Wilson-Fisher).
With local coarse graining, higher-order derivatives and nonlinearities are rendered irrelevant 
by coarse graining.  This then justifies why we only need to renormalize the $\bar{m}^2$ and 
$\bar{\lambda}$ couplings.  However, 
upon non-locally coarse graining the system,
it is no longer possible to ignore the wave-vector dependence that $\lambda$ acquires.
Such perturbations are relevant for $|{\bf p}_i|>1$. 
\begin{table}
\begin{center}
\begin{tabular}{|c|ccc|ccc|}
\hline
& $|p|<1$ & & & $|p|>1$ & & \\
coarse &$p^n\hat{\phi}^2,$ & $p^n\hat{\phi}^m,n>0,$ &  $\hat{\phi}^n,$& 
$p^n\hat{\phi}^2,$ & $p^n\hat{\phi}^m,n>0,$ &  $\hat{\phi}^n,$\\ 
graining&$n>2$& $m \ge 4$ & $n>4$ &$n>2$& $m \ge 4$ & $n>4$ \\
\hline
local & no & no & no & no & no & no \\
nonlocal & no & no & no & yes & yes & yes \\ 
\hline
\end{tabular}
\end{center}
\caption{Relevance of perturbations when D=4}
\label{relevance_table}
\end{table}
Specifically, higher derivative perturbations, 
$p^n\hat{\phi}^2, n>2$ and $p^n\hat{\phi}^m, n>0, m \ge 4$, 
in addition to higher order nonlinearities, $\hat{\phi}^n, n>4$, become relevant when $|{\bf p}_i|>1$.  
While the couplings at small wave-vector, $|{\bf p}_i|<1$, almost obey the standard RG 
equations obtained by local coarse graining, 
if the system is perturbed to its large wave-vector regime, then 
the couplings will flow away from the Gaussian or Wilson-Fisher
fixed points~\cite{shankar94}.  This reflects that the dynamically faster short-wavelength 
perturbations are able to excite the conjugate field $\pi$, 
thereby driving the system away from its standard 
statistical equilibrium.  
Were the conjugate field not accessible to the ``noise'', $\alpha_1=0$, or not an 
observable, $\beta_2=0$, this would not have occurred.
We summarize the relevancy of perturbations in Table \ref{relevance_table} in the case when $D=4$.
A key observation to make from our analysis is that the nonlocal coarse graining produces 
RG equations that are very different from the canonical ones, Equations \eqref{oldRG1}-\eqref{oldRG2}.
In fact, a complete analysis would require us to determine the RG flow of infinitely many coupling constants.

While the existence of an infinity of relevant perturbations renders 
Equations \eqref{RG_eqns1}-\eqref{RG_eqns2b} meaningless, 
for completeness, we present their fixed point structure anyhow.
One 
fixed point of the equations exist where 
$m_<^2 \approx -(4-D)/6$, $\lambda_{(4,0)} \approx 2(4-D)(2\pi)^D/(3S_D)$, $m_>^2 \approx (4-D)/6$, $\lambda_{(0,4)} \approx -2(4-D)(2\pi)^D/(3S_D)$, and the rest
of the couplings zero.  A problem with it is that it doesn't guarantee positivity of the 
action.  This already suggests the need to include higher order, relevant terms.
For instance, to fix the positivity issue, 
we would need to add at least a $\hat{\phi}^6$ term to the action.
Similarly, for $D \le 4$, the irrelevancy of the rest of the $\lambda_{(i,j)}$ terms ensures
that we would lose positivity, thereby requiring the addition of higher order terms.
However, for $D > 4$, there is an alternative fixed-point that guarantees positivity of the action.
At this fixed point, the couplings are  $m_>^2 \approx (4-D)/6$, 
$\lambda_{(0,4)} \approx -2(4-D)(2\pi)^D/(3S_D)$, and the rest zero.
While this fixed point seems strikingly similar to the Wilson-Fisher fixed point, it is important 
to recognize that the coupling constant multiplies products of fields in a
different wave-vector regime.

\section{Conclusion}
\label{conclusion}
In this paper we have presented a new RG procedure and have applied it to a $\phi^4$ toy model.  
We have shown that when both equilibrium and nonequilibrium observables are chosen, 
this RG procedure predicts that na\"ive power counting breaks down and that terms 
that are ordinarily irrelevant become relevant at large wave-vector.
We have shown that the RG equations that we have derived using non-local coarse graining
differ significantly from the RG equations derived local coarse graining.
While our RG equations superficially retain the Gaussian 
and Wilson-Fisher fixed-points as solutions, the infinity of relevant perturbations 
ensure that such fixed points either do not exist or that they pick up an infinity
of unstable directions in the RG flow.
Additionally, the equations superficially admit
many other fixed points.
The generalized Wilsonian RG developed here is applicable to nonequilibrium and 
heterogeneous systems, finite or infinite dimensional systems,
 and systems with various perturbations and uncertainties.
Although the RG is still formally an uncontrolled approximation, the coarse graining is chosen
such that the effective, coarsened system is close to the original one.
Despite the versatility of this method, it is often difficult to analytically determine 
the balancing transformations.  However, since there are very efficient numerical algorithms for 
finding balanced coordinates, this generalized RG remains a numerically useful and practical algorithm.

\acknowledgments
This work was supported by NSF Grant No.~DMR-9813752 and NSF FIBR Grant No.~492153.  
Special thanks are due to Jean Carlson for 
her comments and to Nigel Goldenfeld for his comments and encouragement.

\begin{appendix}

\section{Balancing $X$ and $Y$}
\label{gram_der}

In this appendix we intend 
to calculate the balanced form of the operators $X$ and $Y$
(from Equations \eqref{Lyap1} and \eqref{Lyap2}), called 
gramians, for linear wave equations
(or their discretizations).  This 
entails calculating the damped (exponentially discounted) gramians:
\begin{eqnarray}
\label{eq:damp_gram}
X^{(a)} = \int_0^{\infty} e^{-2at}e^{At}
BB^{\dagger}e^{A^{\dagger}t} \mathrm{d}t  \nonumber \\
Y^{(a)} = \int_0^{\infty} e^{-2at}e^{A^{\dagger}t}
C^{\dagger}Ce^{At} \mathrm{d}t
\end{eqnarray}
As in the body of the paper,
\begin{eqnarray}
B = \left(\begin{array}{cc} \alpha_1\mathbf{I} & 0 \\ 0 & \alpha_2\mathbf{I} \end{array}\right) 
\nonumber \\ 
C = \left(\begin{array}{cc} \beta_1\mathbf{I} & 0 \\ 0 & \beta_2\mathbf{I} \end{array}\right). \nonumber
\end{eqnarray}

Let us first introduce the following notations and conventions.
Recall that any matrix, $\mathbf{S}$, may be expressed in terms of the canonical matrix units,
$\mathbf{e}_{ij}$.  In other words, 
\begin{displaymath}
\mathbf{S} = \sum_{i,j} S_{ij}\mathbf{e}_{ij}
\end{displaymath}
where each $S_{ij}$ is just a complex number.  
For instance, in the case of $2\times2$ matrices,  
\begin{displaymath}
\mathbf{e}_{12} = \left[ \begin{array}{cc} 0 & 1 \\ 0 & 0 \end{array} \right] 
\end{displaymath}
Additionally, for this section, 
$\mathbf{Q} = \left[ \begin{array}{cc} 0 & 1 \\ -1 & 0 \end{array} \right]$.
Lastly, we frequently use the algebraic tensor (dyadic) product, $\otimes$.
For instance, suppose 
\begin{displaymath}
A = \left[ \begin{array}{cc} A_{11} & A_{12} \\ 
A_{21} & A_{22} \end{array} \right]
\end{displaymath}
then
\begin{displaymath}
A\otimes B = \left[ \begin{array}{cc} A_{11}B & A_{12}B \\ 
A_{21}B & A_{22}B \end{array} \right]
\end{displaymath}

First note that if we define 
\begin{equation}
\label{eq:Rmatrix}
\mathbf{R} = \mathbf{e}_{11}\otimes\mathbf{\Omega}^{-1/2} + \mathbf{e}_{22}\otimes\mathbf{\Omega}^{1/2}
\end{equation}
then easily it follows that
\begin{eqnarray}
\label{eq:transform1}
\mathbf{A} = 
 \mathbf{e}_{12}\otimes\mathbf{I} - \mathbf{e}_{21}\otimes\mathbf{\Omega}^2 =
\left[ \begin{array}{cc} 
0 &\mathbf{I} \\  - \mathbf{\Omega}^2 & 0 
\end{array} \right] \nonumber \\
 \stackrel{\mathbf{R}}{\longrightarrow} 
\mathbf{R}^{-1}\mathbf{A}\mathbf{R} = \mathbf{Q}\otimes\mathbf{\Omega}
= \left[ \begin{array}{cc} 
0 &\mathbf{\Omega} \\  - \mathbf{\Omega} & 0 
\end{array} \right]
\end{eqnarray}
From this, one then finds that 
\begin{eqnarray}
\label{eq:trans_cgram}
X^{(a)} = 
\mathbf{R}\int_0^{\infty} e^{-2at}e^{\mathbf{Q}\otimes\mathbf{\Omega}t}
\mathbf{R}^{-1}BB^{\dagger}\mathbf{R}^{-1}e^{-\mathbf{Q}\otimes\mathbf{\Omega}t} \mathrm{d}t\mathbf{R} \nonumber \\
= \mathbf{R}\int_0^{\infty} e^{-2at}e^{\mathbf{Q}\otimes\mathbf{\Omega}t}
\left[ \begin{array}{cc} 
\alpha_1^2\mathbf{\Omega} & 0 \\  0 & \alpha_2^2\mathbf{\Omega}^{-1}
\end{array} \right]
e^{-\mathbf{Q}\otimes\mathbf{\Omega}t} \mathrm{d}t\mathbf{R}
\end{eqnarray}
However, using that $e^{\mathbf{Q}\otimes\mathbf{\Omega}t}$ $= 
\mathbf{I}\otimes\cos\mathbf{\Omega}t + \mathbf{Q}\otimes\sin\mathbf{\Omega}$
we finally arrive at
\begin{eqnarray}
\label{eq:cgram_result}
X^{(a)} = \mathbf{R}\int_0^{\infty} e^{-2at}
\left[\begin{array}{cc}
\alpha_1^2\mathbf{\Omega}\cos^2\mathbf{\Omega}t + \alpha_2^2\mathbf{\Omega}^{-1}\sin^2\mathbf{\Omega}t &
\frac{1}{2}\mathbf{\Omega}^{-1}(\alpha_2^2\mathbf{\Omega}^{-1}-\alpha_1^2\mathbf{\Omega})\frac{\mathrm{d}}{\mathrm{d}t}
\sin^2\mathbf{\Omega}t \\
\frac{1}{2}\mathbf{\Omega}^{-1}(\alpha_2^2\mathbf{\Omega}^{-1}-\alpha_1^2\mathbf{\Omega})\frac{\mathrm{d}}{\mathrm{d}t}
\sin^2\mathbf{\Omega}t &
\alpha_2^2\mathbf{\Omega}^{-1}\cos^2\mathbf{\Omega}t + \alpha_1^2\mathbf{\Omega}\sin^2\mathbf{\Omega}t
\end{array}\right]\mathrm{d}t \ \mathbf{R}
\nonumber \\
= \frac{1}{4a}\mathbf{R}\left[\begin{array}{cc}
\alpha_1^2\mathbf{\Omega} + \alpha_2^2\mathbf{\Omega}^{-1} & 0 \\
0 & \alpha_1^2\mathbf{\Omega} + \alpha_2^2\mathbf{\Omega}^{-1}
\end{array}\right]\mathbf{R} + 
\frac{1}{4a}\mathbf{R}\times \nonumber \\ 
\left[\begin{array}{cc}
-a^2\mathbf{\Omega}^{-1}(\alpha_2^2\mathbf{I}-\alpha_1^2\mathbf{\Omega}^2)
(a^2\mathbf{I}+\mathbf{\Omega}^2)^{-1} &
a(\alpha_2^2\mathbf{I}-\alpha_1^2\mathbf{\Omega}^2)(a^2\mathbf{I}+\mathbf{\Omega}^2)^{-1} \\
a(\alpha_2^2\mathbf{I}-\alpha_1^2\mathbf{\Omega}^2)(a^2\mathbf{I}+\mathbf{\Omega}^2)^{-1} &
 a^2\mathbf{\Omega}^{-1}(\alpha_2^2\mathbf{I}-\alpha_1^2\mathbf{\Omega}^2)
(a^2\mathbf{I}+\mathbf{\Omega}^2)^{-1}
\end{array}\right]\mathbf{R}. 
\end{eqnarray}
Similarly for the other gramian we obtain
\begin{eqnarray}
\label{eq:ogram_result}
Y^{(a)} =
 \frac{1}{4a}\mathbf{R}^{-1}\left[\begin{array}{cc}
\beta_2^2\mathbf{\Omega} + \beta_1^2\mathbf{\Omega}^{-1} & 0 \\
0 & \beta_2^2\mathbf{\Omega} + \beta_1^2\mathbf{\Omega}^{-1} 
\end{array}\right]\mathbf{R}^{-1} + 
 \frac{1}{4a}\mathbf{R}^{-1}\times \nonumber \\ 
\left[\begin{array}{cc}
 a^2\mathbf{\Omega}^{-1}(\beta_1^2\mathbf{I}-\beta_2^2\mathbf{\Omega}^2)
(a^2\mathbf{I}+\mathbf{\Omega}^2)^{-1} &
a(\beta_1^2\mathbf{I}-\beta_2^2\mathbf{\Omega}^2)(a^2\mathbf{I}+\mathbf{\Omega}^2)^{-1} \\
a(\beta_1^2\mathbf{I}-\beta_2^2\mathbf{\Omega}^2)(a^2\mathbf{I}+\mathbf{\Omega}^2)^{-1} &
-a^2\mathbf{\Omega}^{-1}(\beta_1^2\mathbf{I}-\beta_2^2\mathbf{\Omega}^2)
(a^2\mathbf{I}+\mathbf{\Omega}^2)^{-1}
\end{array}\right]\mathbf{R}^{-1}.
\end{eqnarray}

From equations \eqref{eq:cgram_result} and \ref{eq:ogram_result} it 
follows after using $\mathbf{U}_{d}$ to diagonalize $\mathbf{\Omega}$
and taking the small \lq\lq$a$'' limit that the balanced gramian, without ordered eigenvalues,
is given by
\begin{displaymath}
W^{bal} \approx 
 \frac{1}{4a}\left[\begin{array}{cc}
\alpha_1^2\mathbf{\Lambda}_{\Omega} + \alpha_2^2\mathbf{\Lambda}_{\Omega}^{-1} 
& 0 \\
0 & \alpha_1^2\mathbf{\Lambda}_{\Omega} + \alpha_2^2\mathbf{\Lambda}_{\Omega}^{-1} 
\end{array}\right]^{1/2}
\left[\begin{array}{cc}
\beta_2^2\mathbf{\Lambda}_{\Omega} + \beta_1^2\mathbf{\Lambda}_{\Omega}^{-1} 
& 0 \\
0 & \beta_2^2\mathbf{\Lambda}_{\Omega} + \beta_1^2\mathbf{\Lambda}_{\Omega}^{-1} 
\end{array}\right]^{1/2}.
\end{displaymath}

\section{RG equations to 1-loop}
\label{rgeq_derivation}
\input{rgAppendix.tex}

\end{appendix}

\end{document}

%% file: rgAppendix.tex
\begin{fmffile}{supermama}

\newcommand{\twopol}{
\parbox{2cm}{
\begin{fmfchar*}(60,40)
  \fmfleft{i1}
  \fmfright{o1}
  \fmf{plain_arrow}{i1,v}
  \fmf{plain_arrow}{o1,v}
  \fmf{dashes}{v,v}
\end{fmfchar*}
}}

\newcommand{\twopollg}{
\parbox{2cm}{
\begin{fmfchar*}(45,25)
  \fmfleft{i1}
  \fmfright{o1}
  \fmflabel{$\hat{\phi}_>$}{i1}
  \fmflabel{$\hat{\phi}_>$}{o1}
  \fmf{plain_arrow}{i1,v}
  \fmf{plain_arrow}{o1,v}
  \fmf{dashes}{v,v}
\end{fmfchar*}
}}

\newcommand{\twopolll}{
\parbox{2cm}{
\begin{fmfchar*}(45,25)
  \fmfleft{i1}
  \fmfright{o1}
  \fmflabel{$\hat{\phi}_<$}{i1}
  \fmflabel{$\hat{\phi}_<$}{o1}
  \fmf{plain_arrow}{i1,v}
  \fmf{plain_arrow}{o1,v}
  \fmf{dashes}{v,v}
\end{fmfchar*}
}}

\newcommand{\twopb}{
\parbox{2cm}{
\begin{fmfchar*}(60,40)
  \fmfleft{i1}
  \fmfright{o1}
  \fmf{plain_arrow}{i1,v}
  \fmf{plain_arrow}{o1,v}
\end{fmfchar*}
}}

\newcommand{\twopblg}{
\parbox{2cm}{
\begin{fmfchar*}(45,25)
  \fmfleft{i1}
  \fmfright{o1}
  \fmflabel{$\hat{\phi}_>$}{i1}
  \fmflabel{$\hat{\phi}_>$}{o1}
  \fmf{plain_arrow}{i1,v}
  \fmf{plain_arrow}{o1,v}
\end{fmfchar*}
}}

\newcommand{\twopbll}{
\parbox{2cm}{
\begin{fmfchar*}(45,25)
  \fmfleft{i1}
  \fmfright{o1}
  \fmflabel{$\hat{\phi}_<$}{i1}
  \fmflabel{$\hat{\phi}_<$}{o1}
  \fmf{plain_arrow}{i1,v}
  \fmf{plain_arrow}{o1,v}
\end{fmfchar*}
}}

\newcommand{\fourpbllll}{
\parbox{2cm}{
\begin{fmfchar*}(50,30)
  \fmfleft{i1,i2}
  \fmfright{o1,o2}
  \fmflabel{$\hat{\phi}_<$}{i1}
  \fmflabel{$\hat{\phi}_<$}{o1}
  \fmflabel{$\hat{\phi}_<$}{i2}
  \fmflabel{$\hat{\phi}_<$}{o2}
  \fmf{plain_arrow}{i1,v}
  \fmf{plain_arrow}{i2,v}
  \fmf{plain_arrow}{o1,v}
  \fmf{plain_arrow}{o2,v}
\end{fmfchar*}
}}

\newcommand{\fourpblllg}{
\parbox{2cm}{
\begin{fmfchar*}(50,30)
  \fmfleft{i1,i2}
  \fmfright{o1,o2}
  \fmflabel{$\hat{\phi}_<$}{i1}
  \fmflabel{$\hat{\phi}_<$}{o1}
  \fmflabel{$\hat{\phi}_<$}{i2}
  \fmflabel{$\hat{\phi}_>$}{o2}
  \fmf{plain_arrow}{i1,v}
  \fmf{plain_arrow}{i2,v}
  \fmf{plain_arrow}{o1,v}
  \fmf{plain_arrow}{o2,v}
\end{fmfchar*}
}}

\newcommand{\fourpbllgg}{
\parbox{2cm}{
\begin{fmfchar*}(50,30)
  \fmfleft{i1,i2}
  \fmfright{o1,o2}
  \fmflabel{$\hat{\phi}_<$}{i1}
  \fmflabel{$\hat{\phi}_<$}{o1}
  \fmflabel{$\hat{\phi}_>$}{i2}
  \fmflabel{$\hat{\phi}_>$}{o2}
  \fmf{plain_arrow}{i1,v}
  \fmf{plain_arrow}{i2,v}
  \fmf{plain_arrow}{o1,v}
  \fmf{plain_arrow}{o2,v}
\end{fmfchar*}
}}
\newcommand{\fourpblggg}{
\parbox{2cm}{
\begin{fmfchar*}(50,30)
  \fmfleft{i1,i2}
  \fmfright{o1,o2}
  \fmflabel{$\hat{\phi}_<$}{i1}
  \fmflabel{$\hat{\phi}_>$}{o1}
  \fmflabel{$\hat{\phi}_>$}{i2}
  \fmflabel{$\hat{\phi}_>$}{o2}
  \fmf{plain_arrow}{i1,v}
  \fmf{plain_arrow}{i2,v}
  \fmf{plain_arrow}{o1,v}
  \fmf{plain_arrow}{o2,v}
\end{fmfchar*}
}}

\newcommand{\fourpbgggg}{
\parbox{2cm}{
\begin{fmfchar*}(50,30)
  \fmfleft{i1,i2}
  \fmfright{o1,o2}
  \fmflabel{$\hat{\phi}_>$}{i1}
  \fmflabel{$\hat{\phi}_>$}{o1}
  \fmflabel{$\hat{\phi}_>$}{i2}
  \fmflabel{$\hat{\phi}_>$}{o2}
  \fmf{plain_arrow}{i1,v}
  \fmf{plain_arrow}{i2,v}
  \fmf{plain_arrow}{o1,v}
  \fmf{plain_arrow}{o2,v}
\end{fmfchar*}
}}

\newcommand{\fourpb}{
\parbox{2cm}{
\begin{fmfchar*}(60,40)
  \fmfleft{i1,i2}
  \fmfright{o1,o2}
  \fmf{plain_arrow}{i1,v}
  \fmf{plain_arrow}{i2,v}
  \fmf{plain_arrow}{o1,v}
  \fmf{plain_arrow}{o2,v}
\end{fmfchar*}
}}

\newcommand{\fourpol}{
\parbox{2cm}{
\begin{fmfchar*}(60,40)
  \fmfleft{i1,i2}
  \fmfright{o1,o2}
  \fmf{plain_arrow}{i1,v1}
  \fmf{plain_arrow}{i2,v1}
  \fmf{plain_arrow}{o1,v2}
  \fmf{plain_arrow}{o2,v2}
  \fmf{dashes, left}{v1,v2,v1}
\end{fmfchar*}
}}

\newcommand{\fourpolllll}{
\parbox{2cm}{
\begin{fmfchar*}(50,30)
  \fmfleft{i1,i2}
  \fmfright{o1,o2}
  \fmflabel{$\hat{\phi}_<$}{i1}
  \fmflabel{$\hat{\phi}_<$}{o1}
  \fmflabel{$\hat{\phi}_<$}{i2}
  \fmflabel{$\hat{\phi}_<$}{o2}
  \fmf{plain_arrow}{i1,v1}
  \fmf{plain_arrow}{i2,v1}
  \fmf{plain_arrow}{o1,v2}
  \fmf{plain_arrow}{o2,v2}
  \fmf{dashes, left}{v1,v2,v1}
\end{fmfchar*}
}}

\newcommand{\sixpb}{
\parbox{2cm}{
\begin{fmfchar*}(60,40)
  \fmfleft{i1,i2,i3}
  \fmfright{o1,o2,o3}
  \fmf{plain_arrow}{i1,v1}
  \fmf{plain_arrow}{i2,v1}
  \fmf{plain_arrow}{i3,v1}
  \fmf{plain_arrow}{o1,v2}
  \fmf{plain_arrow}{o2,v2}
  \fmf{plain_arrow}{o3,v2}
  \fmf{dashes}{v1,v2}
\end{fmfchar*}
}}

In this Appendix, we briefly derive the RG equations for the 
two-point (Eqs. \eqref{RG_eqns2a} and \eqref{RG_eqns2b}), and the 
four-point coupling functions (Eq. \eqref{RG_eqns1}).
In  Section \ref{sec:pert_theory} we sketch out the standard perturbative procedure 
\cite{wilson74,wilson75,shankar94,goldenfeld92} used to derive the RG equations.
Then we proceed to use the perturbative procedure to derive the RG equations for 
the two-point and four-point couplings in Sections \ref{sec:2point} and \ref{sec:4point}
respectively.

\subsection{Sketch of perturbation theory}
\label{sec:pert_theory}

Recall that  we will be integrating out the field variables $\hat{\phi}_m$ and 
keeping $\hat{\phi}_<$ and $\hat{\phi}_>$.  
Ideally we would like to exactly evaluate the partial trace 
\begin{equation}
\label{eq:partial_trace1}
\int \mathscr{D}\hat{\phi}_m \exp\left\{-S(\hat{\phi}_<, \hat{\phi}_>,\hat{\phi}_m)\right\} = 
\exp\left\{-\bar{S}(\hat{\phi}_<, \hat{\phi}_>,)\right\}.
\end{equation}
In general, computing such a functional integral is very difficult.  Consequently, 
we will perturbatively evaluate Equation \eqref{eq:partial_trace1} about the quadratic 
part of the action.
Now let us denote the quadratic 
part of the action by $S_2$ and the remainder by $S_r$ (i.e.~$S=S_2+S_r$).  By the simple form 
of the quadratic part of the action in Equation \eqref{action}, we have that
\begin{equation}
\label{quad_decomp}
S_2(\hat{\phi}_<, \hat{\phi}_>,\hat{\phi}_m) = S_2(\hat{\phi}_<, \hat{\phi}_>) + S_2(\hat{\phi}_m).
\end{equation}
Now given that $\mathcal{P}_0\{\hat{\phi}_m\} = Z_{0,m}^{-1}\exp\{-S_2(\hat{\phi}_m)\}$ is
the probability distribution generated by $S_2(\hat{\phi}_m)$, let $\langle \hdots \rangle_0$
denote averages taken with respect to this distribution.
Using this notation, Equation \eqref{eq:partial_trace1} becomes
\begin{equation}
\label{eq:partial_trace2}
\int \mathscr{D}\hat{\phi}_m \exp\left\{-S(\hat{\phi}_<, \hat{\phi}_>,\hat{\phi}_m)\right\} = 
\exp\left\{-S_2(\hat{\phi}_<, \hat{\phi}_>)\right\}\left<\exp\left\{-S_r\right\}\right>_0.
\end{equation}
Note that we have absorbed any contribution from $Z_{0,m}$ into the path integral measure.

Equation \eqref{eq:partial_trace2} may be (approximately) calculated by expansion techniques.
Two techniques that are frequently used are the cummulant expansion \cite{shankar94,goldenfeld92}
and Feynman diagram expansions \cite{wilson74}.  Although Feynman diagram expansion methods permit 
infinite resummations more easily than the cummulant expansion, we make use of the cummulant 
expansion.  However, we will still illustrate the non-vanishing Feynman diagrams that contribute 
to the cummulant expansion.  Upon applying the cummulant expansion 
to Equation \eqref{eq:partial_trace2} we find that
\begin{equation}
\label{eq:cum_expans}
\left<\exp\left\{-S_r\right\}\right>_0 = \exp\left\{-\left<S_r\right>_0 
+ \frac{1}{2}\left[\left<S_r^2\right>_0-\left<S_r\right>_0^2\right] - \hdots \right\}.
\end{equation}
The first term in the exponent on the right hand side of Equation \eqref{eq:cum_expans} 
is the first cummulant while the second term (in square brackets)
is the second cummulant.  
If we take $S'$ to be the exponent on the right hand side of Equation \eqref{eq:cum_expans}
then the Feynman diagram representation of $S'$ is given by
\begin{eqnarray}
\label{feynman_exp}
S' = & \twopb\  + \ 6 \ \twopol \ + \ \fourpb \nonumber \\
&+ \ 6\cdot 6\fourpol \ + \ \sixpb \ + \ \hdots,
\end{eqnarray}
where the solid propagators represent the $\hat{\phi}_<$ and $\hat{\phi}_>$ propagators
while the dashed propagator represent the $\hat{\phi}_m$ propagator.
We will use the first and second cummulants (and diagrammatics) to derive the RG equations 
in the following sections.

\subsection{Two-point couplings}
\label{sec:2point}

If we consider having a mass term (i.e.~$\nabla^2 \to \nabla^2-m^2$), then 
after integrating out $\hat{\phi}_m$ 
and rescaling $\hat{\phi}_>$ and $\hat{\phi}_<$ according Equation \eqref{rescale}, 
the bare two-point propagator (Green's function) becomes 
\begin{eqnarray}
\label{2pt_tree}
\twopb \ = \ & \  \twopbll \ \ \ + \ \ \ \ \ \twopblg \nonumber \\
     \  = \ & \frac{1}{2}\left(\int_q q^2|\hat{\varphi}(q)|^2 +  \Lambda^{-2}\int_{|q|<1}m^2_<|\hat{\varphi}(q)|^2
+ \Lambda^{2}\int_{|q|>1}m^2_>|\hat{\varphi}(q)|^2\right).
\end{eqnarray}
To simplify notation we denote the free propagator by  $G_0(p)$.  Similarly, for
$|p|<1$ we denote the propagator by  $G_0^<(p)$  and we do similarly for $|p|>1$.
At one loop we obtain
\begin{eqnarray}
\label{2pt_1loop}
\twopol \ = \ &\  \twopolll \ \ \ + \ \ \ \ \ \twopollg \nonumber \\
       = & \frac{1}{4!}\left( \Lambda^{-2}\left[\lambda_{(4,0)}\int_{\Lambda<|p|<1} G_0^<(p)
+ \lambda_{(2,2)}\int_{1<|p|<\lambda^{-1}} G_0^>(p)\right]\int_{|q|<1}|\hat{\varphi}(q)|^2  \right. \nonumber \\
 & \left. +  \Lambda^{2}\left[\lambda_{(2,2)}\int_{\Lambda<|p|<1} G_0^<(p)
+ \lambda_{(0,4)}\int_{1<|p|<\lambda^{-1}} G_0^>(p)\right]\int_{|q|>1}|\hat{\varphi}(q)|^2\right) \\
= & \frac{\lambda S_D}{4!(2\pi)^D}\left(1-\Lambda\right)\left(\left[\frac{\lambda_{(4,0)}}{m^2_<+1}+\frac{\lambda_{(2,2)}}{m^2_>+1}\right]\int_{|q|<1}|\hat{\varphi}(q)|^2 + \left[\frac{\lambda_{(2,2)}}{m^2_<+1}+\frac{\lambda_{(0,4)}}{m^2_>+1}\right]\int_{|q|>1}|\hat{\varphi}(q)|^2\right)
\end{eqnarray}
where $G_0(p)$ is the free propagator and $S_D$ is the area of the $D$-dimensional unit sphere.

By combining Equations \eqref{eq:cum_expans}, \eqref{feynman_exp}, \eqref{2pt_tree}, and \eqref{2pt_1loop} and using 
$\Lambda = e^{-dl}$, we arrive at the RG equations (Eqs. \eqref{RG_eqns2a} and \eqref{RG_eqns2b})
\begin{eqnarray}
\partial_lm^2_< = 2m^2_< + \frac{S_D}{2(2\pi)^{D}}\left(\frac{\lambda_{(4,0)}}{m_<^2+1}+\frac{\lambda_{(2,2)}}{m_>^2+1}\right)+\mathscr{O}({\lambda^2}) \\
\partial_lm^2_> = -2m^2_> + \frac{S_D}{2(2\pi)^{D}}\left(\frac{\lambda_{(2,2)}}{m_<^2+1} +\frac{\lambda_{(0,4)}}{m_>^2+1}\right)+\mathscr{O}({\lambda^2})
\end{eqnarray}

\subsection{Four-point couplings}
\label{sec:4point}

The decoupling of the wave-vector regimes, while explicit in the calculation of the 2-point 
couplings, is more subtle in the calculation of the 4-point couplings.  However, it is 
still manifestly evident at tree level.  Consequently, as in the previous section, we will 
first calculate the rescalings for the bare 4-point vertex.

\begin{eqnarray}
\label{4ptb4_0}
\fourpbllll \ \ = & \ \Lambda^{D-4} \frac{\lambda_{(4,0)}}{4!(2\pi)^{4D}}\int_{|p_i|<1}\prod_i
\frac{d^Dp_i}{(2\pi)^D} 
(2\pi)^D \nonumber \\
& \delta^D\left(\sum_{j=1}^4 p_j\right)\varphi(p_1)\varphi(p_2)\varphi(p_3)\varphi(p_4),
\end{eqnarray}
\begin{eqnarray}
\label{4ptb3_1}
\fourpblllg \ \ = & \ \Lambda^{2(D-1)} \frac{\lambda_{(3,1)}}{4!(2\pi)^{4D}}\int_{|p_i|<1}\prod_{i=1}^3
\frac{d^Dp_i}{(2\pi)^D} \int_{|p_4|>1} \frac{d^Dp_4}{(2\pi)^D} (2\pi)^D \nonumber \\
& \delta^D\left(\sum_{j=1}^4 p_j\right)\varphi(p_1)\varphi(p_2)\varphi(p_3)\varphi(p_4),
\end{eqnarray}
\begin{eqnarray}
\label{4ptb2_2}
\fourpbllgg \ \ = & \ \Lambda^{D} \frac{\lambda_{(2,2)}}{4!(2\pi)^{4D}}\int_{|p_i|<1}\prod_{i=1}^2
\frac{d^Dp_i}{(2\pi)^D} \int_{|p_k|>1} \prod_{k=3}^4 \frac{d^Dp_k}{(2\pi)^D} (2\pi)^D \nonumber \\
& \delta^D\left(\sum_{j=1}^4 p_j\right)\varphi(p_1)\varphi(p_2)\varphi(p_3)\varphi(p_4),
\end{eqnarray}
\begin{eqnarray}
\label{4ptb1_3}
\fourpblggg \ \ = & \ \Lambda^{2} \frac{\lambda_{(1,3)}}{4!(2\pi)^{4D}}\int_{|p_i|>1}\prod_{i=2}^4
\frac{d^Dp_i}{(2\pi)^D} \int_{|p_1|<1} \frac{d^Dp_1}{(2\pi)^D} (2\pi)^D \nonumber \\
& \delta^D\left(\sum_{j=1}^4 p_j\right)\varphi(p_1)\varphi(p_2)\varphi(p_3)\varphi(p_4),
\end{eqnarray}
\begin{eqnarray}
\label{4ptb0_4}
\fourpbgggg \ \ = & \ \Lambda^{4-D} \frac{\lambda_{(0,4)}}{4!(2\pi)^{4D}}\int_{|p_i|>1}\prod_i
\frac{d^Dp_i}{(2\pi)^D}
(2\pi)^D \nonumber \\
& \delta^D\left(\sum_{j=1}^4 p_j\right)\varphi(p_1)\varphi(p_2)\varphi(p_3)\varphi(p_4).
\end{eqnarray}

At one loop, a sample calculation with only $\hat{\phi}_<$ on the external legs 
of the 4-point diagram produces
\vspace{.2cm}
\begin{eqnarray}
\label{4ptol4_0a}
\fourpolllll \ \ = & \ \left(\frac{1}{4!}\right)^2
(2\pi)^{-4D}\int_{|q_i|<\Lambda}\int_{\Lambda<|Q_i|<\Lambda^{-1}}\lambda(q_1,q_2,Q_1,Q_2)
(2\pi)^D\delta^D\left( q_1+q_2+Q_1+Q_2\right) \nonumber \\
& \lambda(q_3,q_4,Q_3,Q_4)(2\pi)^D\delta^D\left(q_3+q_4+Q_3+Q_4\right)
(2\pi)^D\delta^D\left(Q_1+Q_3\right)G_0(Q_1) \nonumber \\
& (2\pi)^D\delta^D\left(Q_2+Q_4\right)G_0(Q_2) \varphi(q_1)\varphi(q_2)\varphi(q_3)\varphi(q_4) 
\end{eqnarray}
\begin{eqnarray}
\label{4ptol4_0b}
\ \approx & \ \left(\frac{1}{4!(2\pi)^{D/2}}\right)^2\int_{|q_i|<1}\left[
\int_{\Lambda \le |Q_1|,|Q_2| < 1}\lambda_{(4,0)}^2G_0^<(Q_1)G_0^<(Q_2) \right. \nonumber \\
 & +2\int_{\Lambda \le |Q_1| < 1}\int_{1 \le |Q_2| \le \Lambda^{-1}}\lambda_{(3,1)}^2G_0^<(Q_1)G_0^>(Q_2)
\nonumber \\
& \left.+\int_{1 \le |Q_1|,|Q_2| \le 1}\lambda_{(2,2)}^2G_0^>(Q_1)G_0^>(Q_2)\right] \nonumber \\
& \delta^D\left(q_1+q_2+Q_1+Q_2\right)(2\pi)^D\delta^D\left(q_3+q_4-Q_1-Q_2\right)
\varphi(q_1)\varphi(q_2)\varphi(q_3)\varphi(q_4).
\end{eqnarray}
From equation \eqref{4ptol4_0b}, there is an acquired wave-vector dependence. 
For the sake of comparison with equations, we only wish to derive the (naive) 
flow equations for the constant part of $\lambda(q_1,q_2,q_3,q_4)$. Thus, we obtain

\vspace{.2cm}
\begin{eqnarray}
\label{4ptol}
\fourpolllll  \ \  \approx & \ S_D(1-\Lambda)\left(\frac{1}{4!(2\pi)^{D/2}}\right)^2
\left[\lambda_{(4,0)}^2\left(1+m^2_<\right)^{-2} \right. \nonumber \\
& + 2\lambda_{(3,1)}^2\left(1+m^2_<\right)^{-1}\left(1+m^2_>\right)^{-1} 
\left.+ \lambda_{(2,2)}^2\left(1+m^2_>\right)^{-2}
\right]
\nonumber \\
 & \int_{|q_i|<1}(2\pi)^D
\delta^D\left(\sum_{j=1}^4 q_j\right)\varphi(q_1)\varphi(q_2)\varphi(q_3)\varphi(q_4)
\end{eqnarray}
Combining Equations \eqref{eq:cum_expans}, \eqref{feynman_exp}, 
\eqref{4ptb4_0}-\eqref{4ptb0_4}
, and \eqref{4ptol} 
and using
$\Lambda = e^{-dl}$ yields the RG equations given in Equation \eqref{RG_eqns1}.

\end{fmffile}